\documentclass[aps,prl,twocolumn,showpacs,amsmath,amssymb]{revtex4}
\usepackage{graphicx}

\def\e{{\mathrm e}}
\def\dd{{\mathrm d}}
\def\lr#1#2#3{\left#1{#3}\right#2}

\begin{document}
\title{Low-Temperature Decoherence of Qubit Coupled to Background Charges}

\author{Alex Grishin, Igor V.\ Yurkevich, and Igor V.\ Lerner}
\affiliation{School of Physics and Astronomy, University of
Birmingham, Birmingham B15 2TT, United Kingdom}

\date{December 14, 2004}
\begin{abstract}
We have found an exact expression for the decoherence rate of a
Josephson charge qubit coupled to fluctuating background charges.
At low temperatures $T$ the decoherence rate ${\Gamma}$ is linear
in $T$ while at high temperatures it saturates in agreement with a
known classical solution which, however,  reached  at surprisingly
high $T$. In contrast to the classical picture, impurity states
spread in a wide interval of energies ($\gg T$) may essentially
contribute to ${\Gamma}$.

\end{abstract}
\pacs{73.21.-b, 74.78.Na, 03.65.Yz  }

\maketitle

Solid state nanoscale devices provide one of the most promising
routes to implementing a scalable set of controllable two-state
quantum systems, qubits, based either on  spin degrees of freedom
of electrons in quantum dots \cite{LossDiV} or on discrete charge
quantum states in Josephson junctions \cite{shnirman}. An
unavoidable coupling of each qubit to the environment leads to
decoherence. The loss of coherence before a sufficient amount of
quantum operations was performed would be the major impediment in
using solid-state qubits in quantum computations.

Recent experiments on Josephson-junction (JJ) single qubits
\cite{nakamura3,vion}  have demonstrated the possibility of
performing hundreds of ``quantum operations''  (i.e.\ coherent
oscillations between the qubit states) before environmental
decoherence sets in. It is believed, however, that tens of
thousands of such operations are required for quantum computation
to become a reality \cite{LossDiV2} so that much longer
decoherence times should be achieved experimentally. It
necessitates a better theoretical understanding of realistic
mechanisms of decoherence.

The most conventional way to describe environmental decoherence is
based on the spin-boson models (see for reviews \cite{leggett})
where the qubit interacts with environment represented as a set of
harmonic oscillators with a given frequency spectrum. An
alternative approach is based on identifying the dominant
mechanism of decoherence in a real experimental situation and
formulating the adequate model. The results obtained in such a
model are not necessarily reducible to those in the spin-boson
models. This turns out to be the case for decoherence in  charge
JJ qubits addressed in this Letter.

It is widely believed that in  charge  qubits \cite{nakamura3} the
main contribution to decoherence  comes from an inevitable
coupling to charge degrees of freedom, which is also responsible
for $1/f$ noise observed in such devices \cite{zorin,nakamura}. As
charge impurities are spatially frozen as  experimental
temperatures $T\sim 30\div50$mK \cite{nakamura3,zorin,nakamura},
the most probable source of dynamical electromagnetic fields is
impurities recharging, e.g.\ due to the hybridization of their
electronic states   with conduction electrons (in the metallic
electrodes, etc.) The appropriate model similar to the
conventional model of the spectral diffusion in glasses
\cite{anderson} is known as the spin-fluctuator model. It has
already been used \cite{paladino,galperin,paladino2,faoro} for a
classical (``high-temperature'') description of decoherence and
noise in  charge JJ qubits. However, in such a description
contributions from impurities with energy levels outside a narrow
(of order $T$) strip around the Fermi level of conduction
electrons are exponentially suppressed. Had this been the case,
decoherence from fluctuating charge impurities would hardly be
seen in experiments as it is rather unlikely to find such a
fluctuator coupled to the qubit in the energy strip so narrow
compared to the typical (atomic) scale over which impurity levels
are distributed.

In this Letter we develop a complete quantum mechanical
description of decoherence due to charge fluctuators and obtain
\textit{asymptotically exact} expression for the decoherence rate
$\Gamma(T)$ given by Eq.~(\ref{finiteT}). In the classical
``high-$T$'' regime it goes over to the previously obtained result
\cite{paladino,galperin} while in the low-$T$ regime $\Gamma(T)$
{\it decreases linearly} with $T$, Eq.~(\ref{lowT}), having a
nontrivial non-monotonic dependence on the coupling strength $g$.
In particular,  the exponential suppression of the contributions
of energetically remote impurities turns out to be an artefact of
the classical description. The hybridization with the conduction
electrons responsible for the dynamical recharging leads also to a
quantum broadening of the impurity levels which results in the
contribution of (inevitably present) energetically remote
impurities being suppressed only as power-law rather than
exponentially, thus making it detectable and eventually dominant.

We consider a model where a charge qubit is coupled to impurities
with charge states fluctuating due to hybridization with the
conduction band:
\begin{align}
&\hat{H}=\frac{\omega_0}{2}\hat{\sigma}_z-\frac{E_J}{2}
\sigma_x+\hat{\sigma}_z\,\hat{V}+\hat{H}_B,\quad \hat{V}\equiv
\frac{1}{2}\sum_i\,v_i\,\hat{d}^{\dagger}_i\hat
d^{\phantom{\dagger}}_i
\notag\\[-6pt]\label{htotal}
\\[-6pt]\notag
&\hat{H}_{B}=\sum\limits_{i}\varepsilon^{0}
_{i}\hat{d}^{\dagger}_i\hat d^{\phantom{\dagger}}_i+
\sum\limits_{i,\mathbf{k}}\left[t^{\phantom{\dagger}}
_{\mathbf{k}i}\hat{c}^{\dagger}_\mathbf{k}\hat
d^{\phantom{\dagger}}_i +\text{h.c.}\right]+
\sum\limits_{\mathbf{k}}\varepsilon^{\phantom{\dagger}}_{\mathbf{k}}
\hat{c}^{\dagger}_{\mathbf{k}}\hat
c^{\phantom{\dagger}}_{\mathbf{k}}\,.
\end{align}
Here $\omega_0$  is the  energy split and  $E_J$ is the Josephson
coupling of the two qubit states separated from the higher states
by the Coulomb blockade energy $E_c\gg T$; both $\omega_0$    and
$E_J$ are tuneable which allows one to perform quantum operations
on the qubit. Each localized impurity state is characterized by
its coupling to the qubit $v_i$ (due to the dipole interaction),
its energy ${\varepsilon} _i$ (counted from the Fermi energy
${\varepsilon} _F$ of the conduction electrons), and its
switching rate  ${\gamma} _i=2\pi \nu |t_i|^2$ ($\nu$ is the
density of states at  ${\varepsilon} _F$  in the conduction band
and  $|t_i|^2=\operatorname{Vol}\cdot|t_{i\mathbf k}|^2$). All the
 parameters  ${\varepsilon} _i^0,\; {\gamma} _i$ and
$v_i$,  vary independently   in a wide interval of energies.

Solving the Heisenberg equation of motion for the full density
matrix with  separable initial conditions,
$\hat{\rho}_{\text{B+Q}}(0)=\hat{\rho}
(0)\otimes\hat{\rho}_{\text{B}} $, where $\hat\rho(t)$ is the
reduced density matrix of the qubit and
$\hat\rho_{\text{B}}=Z_{\text{B}}^{-1}\e^{-{\beta} \hat
H_{\text{B}}}$ is the equilibrium density matrix of the bath, one
writes $\hat\rho(t)$ in the standard formal representation (see,
e.g., \cite{galperin}):
\begin{align}
\hat{\rho} (t)
= \begin{pmatrix}  n(t) & \rho _{12}(0)\e^{-i\omega_0t}D(t)\\
\rho _{21}(0)\e^{ i\omega_0t}D^*(t) & 1-n(t)\end{pmatrix}
\label{reduced}
\end{align}

For the charge qubit under consideration, the operational
condition is $E_J\ll{\omega} _0$ \cite{nakamura3}. We are
interested in decoherence only so that we restrict considerations
to the `pure dephasing' regime \cite{paladino}, $E_J=0$.  In the
perturbative region, small $E_J$ would not lead to noticeable
corrections to decoherence. Although  $E_J$ could be tuned to a
large value if an  operation is performed on the qubit, this
should happen  only for a short period of time  which  gives only
negligible corrections to decoherence.

For $E_J=0$ the coupling ${\sigma} _z\hat V$  the diagonal
elements of $\hat \rho$ do not evolve while the time evolution of
the off-diagonal elements of   $\hat \rho$ can be represented as
\begin{align}\label{f}
D(t )=\left\langle \e^{i(\hat{H}_B+\hat{V})t
}\,\e^{-i(\hat{H}_B-\hat{V})t }\right\rangle_{\text{B}}
\end{align}
where $\langle\ldots\rangle_{{\text{B}}}$ is the average with the
Gibbs density matrix of the bath, $\hat \rho_{\text{B}}$.
 Expressions of this sort can be exactly calculated in certain problems,
e.g.\ the orthogonality catastrophe \cite{mahan} and full counting
statistics \cite{Levitov} in $1D$, in techniques which are always
problem-specific. In present considerations, we employ the
linked-cluster expansion within the Keldysh formalism (similar to
that used in  \cite{shnirman3} for bosonic environment) to find
the decoherence rate ${\Gamma}$ defined by
\begin{align}
\Gamma(T)=-\lim_{t \to\infty}\,\,t ^{-1}\ln\left|D(t
)\right|\,.\label{dec}
\end{align}
In order to calculate ${\Gamma}$, we represent $D(t)$ in
Eq.~(\ref{f}) as the following functional integral over the
Grassmann fields defined on the standard Keldysh contour
$c_{_\text{K}}$ \cite{rammer}:
\begin{align}\notag
D(t)&=\int\!\frac{\mathcal{D}\bar d\mathcal{D}d\mathcal{D}\bar c
\mathcal{D}c}{ \mathcal{Z}}  \exp\biggl[{ i\! \int
_{c_{_\text{K}}}\!\!\!\! \dd
t'\biggl({\sum\limits_{ij}S_{ij}+\sum\limits_{\mathbf k}S_{\mathbf
k} }\biggr)\biggr]}\,,
\end{align}
where the action densities are given by
 \begin{align}\notag S_{ij}&=
\bar d _j(t')\left(i\partial_{t'}-\varepsilon_{j}^0
+\frac{v_j(t')}{2}\right)d_j(t'){\delta}_{ij}\,,
 \\[-6pt]\label{action}\\
S_{\mathbf k} &= \bar c _{\mathbf k}(t')(i\partial_{t'}\! -\!
\varepsilon_{\mathbf k})c_{\mathbf k}(t') -\sum\limits_{i
}\Big(t_{\mathbf ki}\bar c _{\mathbf
k}(t')d_i(t')+\text{h.c.}\Big). \notag
\end{align}
The  field  $v_j(t')=\pm v_j$
 for $0\leq t' \leq t $, with `$+$' sign on the upper and `$-$' sign on
 the lower branch of the
Keldysh contour, and vanishes for $t'<0$ or $t'>t$. The
normalization $\mathcal{Z}$ is defined as the same functional
integral but with $v_j\equiv 0$. The integration over the
conduction electron fields $\bar  c _{\mathbf k},\, c _{\mathbf k}
$ reduces the action (\ref{action})  to the impurity term $S_{ij}$
with the mass operator
\begin{align}
\Sigma^{\phantom{*}}_{ij}(t',t'')=\sum_{\mathbf
k}t^{\phantom{*}}_{\mathbf {k}i}{ t}^*_{\mathbf
{k}j}g^{\phantom{*}}_{\mathbf  k}(t',t'')\,,\label{mass}
\end{align}
where $g_{\mathbf k}(t',t'')$ is the conduction electron Green
function obeying the equation $\lr(){i\partial _{t'}
-{\varepsilon} _{{\mathbf k}}} g_{\mathbf
k}(t',t'')={\delta}(t',t'')$  with the delta function defined on
the Keldysh contour.
 Now the integration over the fields $\bar d _{i}$ and $ d _{i} $
reduces $D(t)$,  Eq.~(\ref{f}),  to the appropriate matrix
determinant thus yielding the following formal result for
${\Gamma}$, Eq.~(\ref{dec}):
\begin{align}\label{Gamma}
\Gamma(T)=-\Re{\e}\lim_{t \to\infty}\,\,t
^{-1}\operatorname{Tr}\ln\left[1+\frac{\hat{v}}{2}\hat{G}\right]
\end{align}
where $\hat G$ obeys  $ \lr(){i\partial _{t'} -\hat{\varepsilon}^0
-\widehat {\Sigma} } \hat G=\hat I\,, $ and $\operatorname{Tr}$
implies an integration over times along the Keldysh contour and a
summation over indices labelling the fluctuators.  The dependence
on the running time $t$ above is via $v_j(t)$ defined after
Eq.~(\ref{action}). The long-$t$ limit  in Eq.~(\ref{Gamma}) can
be found by expanding $\operatorname{Tr}\ln$ into power series in
$\hat v\hat G$. The $n^{{\text{th}}}$  order term of the expansion
is a multiple integral over $n$ time variables, each  running
along that part of $c_{_\text{K}}$ where $v\ne0$. As usual
\cite{rammer}, we represent each integrand  via the Keldysh
matrices
\begin{align}\label{keldysh}
    \hat {\mathcal{G}}=\begin{pmatrix}
      \hat G^R & \hat G^K  \\
      0 & \hat G^A\\
    \end{pmatrix} \begin{matrix}
        &   \\
      , &  \\
    \end{matrix}
\end{align}
thus reducing each  integration over time to that from $0$ to $t$
(with $v\to v{\tau} _x$, where ${\tau} _x$ is the Pauli matrix in
the Keldysh space). Due to the time translation invariance, each
$\hat G$ depends only on the difference of its time arguments.
Then the $n^{{\text{th}}}$ order integrand depends on $n-1$
differences in times, while the integration over the last time
variable produces the overall factor of $t$. The remaining
integrals in time can be extended to the entire axis as all the
Green functions exponentially decay in time with the time
constants ${\gamma} ^{-1}_i$ (or $T^{-1}$ for the real part of the
Keldysh component $G^{\text{K}}$). Then the integral has a
convolution structure in time and, upon a Fourier transform, it
finally reduces to the
 integral of $t\lr[]{\frac{1}{2 } \hat v
\mathcal{G}({\omega} )}^n $ over ${\omega} $. Since the coefficients of the
expansion were not affected by the Fourier transform, the
re-summing the series restores the logarithm and Eq.~(\ref{Gamma})
reduces to
\begin{align}
\Gamma=-\Re{\e}\int\limits_{-\infty}^{+\infty}\frac{\dd\omega}{2\pi}
\operatorname{tr}\ln \left[1+\frac{ {v}}{2}\hat G^{\text K}-\frac{
{v}}{2}\hat{G}^{\text R}\frac{ {v}}{2}\hat{G}^{\text
A}\right]\label{Det}
\end{align}
where the trace of $\ln\Big[1+ \frac{\hat v}{2} {\tau} _x \hat
{\mathcal{G}}({\omega} )\Big]$ in the space of matrices
(\ref{keldysh})   has been explicitly taken and
$\operatorname{tr}$ refers only to the fluctuator matrix indices.

It follows from Eq.~(\ref{Det}) that $\Gamma_0\!\equiv
\!\Gamma(T\!=\!0)$ vanishes as expected. Indeed, at $T\!=\!0$ one
uses \cite{rammer} $G^{\text K}(\omega)=\left[G^{\text
R}(\omega)-G^{\text A}(\omega)\right]\operatorname{sgn}\omega$ to
find
\begin{align}\label{zero}
\Gamma_0=-\Re{e}\!\!\!\int\limits_{-\infty}^{+\infty}\!\frac{\dd\omega}{2\pi}
\operatorname{tr}\ln\! \Big[\! \Big(1+\frac{\hat{v}}{2}\hat
G^{\text R}\Big)\!\!\Big(1-\frac{\hat{v}}{2}\hat G^{\text
A}\Big)\! \Big]\!=0\,,
\end{align}
since the first order of the expansion is imaginary, while the higher orders vanish upon the integration as all
the poles are in the upper (lower) ${\omega} $ half-plane.

A further simplification is possible for a typical situation when
distances between the fluctuators are larger than the Fermi
wavelength $k_{\text{F}}^{-1}$ so that the hybridization is local,
$ t_{\mathbf ki}= t_i\,V^{-1/2}\e^{i\mathbf {k}\mathbf {r}_i} $.
Then Eq.~(\ref{mass}) reduces to $\Sigma_{ij}=
t^{\phantom{*}}_it^*_jg(\mathbf {r}_i\!-\!\mathbf {r}_j, t'\!-\!
t'')$. Since $g(\mathbf {r})$ (the Fourier transform of the
conduction electron Green function $g_{\mathbf k})$ oscillates at
$k_{\text{F}}^{-1}$, off-diagonal matrix elements of $\widehat
{\Sigma} $ vanish upon the integration (neglecting  small
interference corrections) making $\hat G$ in Eq.~(\ref{Gamma})
diagonal in the fluctuator indices, $\hat G=G_{j}{\delta}_{ij}$
(such a diagonal form was assumed in \cite{paladino} by choosing
independent conduction bands for each impurity as in
\cite{nazarov}.) Here
 \begin{align}
    &G^{\text  R/A}_{ j}(\omega)
= \left(\omega-\varepsilon_{j}\pm i\frac{\gamma_j}2\right)^{-1},
 \label{GR}
\end{align}
where ${\gamma} _j\equiv 2\Im{\text{m}} {\Sigma}
^{\text{R}}_{jj}=2\pi \nu |t_j|^2$ and ${\varepsilon}
_j={\varepsilon} _j^0+\Re{\text{e}}{\Sigma} _{jj}$ is  the
fluctuator energy (counted from ${\varepsilon}_{\text{F}} $)
renormalized by the hybridization. The Keldysh component  of
$\hat{\mathcal{G}}$ is given by \cite{rammer} $G^{\text K}
(\omega)= [G^{\text R} (\omega)-G^{\text A}_{j}(\omega) ] [1-2n_{
_{\text F}}(\omega)]$, where $n_{ _{\text F}}(\omega)$ is the
Fermi factor with ${\omega} $ counted from ${\varepsilon}
 _{ _{\text F}} $.

Then ${\Gamma}$ in Eq.~(\ref{dec}) reduces to a sum of the
individual fluctuator  contributions, $
\Gamma(T)=\sum_j\,\Gamma_j(T)$. Substituting  $\hat G_j$ of
Eq.~(\ref{GR}) into Eq.~(\ref{Det}), subtracting the identically
zero expression (\ref{zero}) for ${\Gamma}_0$ (to improve the
integral convergency) and taking the real part of the resulting
expression, we obtain the following contribution of a single
fluctuator at energy ${\varepsilon} _j \equiv {\varepsilon}$ to
the decoherence rate:
\begin{align}
   \Gamma_{\varepsilon} (T)=-\int\limits_{-\infty}^{+\infty}\frac{d\omega}{4\pi}
\ln\Bigg\{ 1-\frac{4n_{ _{\text F}}(\omega) \lr[]{1-n_{ _{\text
F}}(\omega)}}{1+\left[ \lambda^{-1}_{\varepsilon} (\omega) -\frac12g
\right]^2}\Bigg\}\,.\label{finiteT}
\end{align}
Here we  suppressed the index $j$ and introduced dimensionless
coupling $g\equiv v/{\gamma} $  of the qubit and the  fluctuator
 with the dimensionless
density of states ${\lambda}_{\varepsilon} (\omega) \equiv \pi v
\nu_{\varepsilon} ({\omega} )$ broadened  around  the energy $
{\varepsilon}$ by the hybridization:
 \begin{align}
\label{nu} \nu_{\varepsilon}(\omega)
=-\frac{1}{\pi}\Im{\text{m}}\, G^{\text
R}_{\varepsilon}(\omega)&=\frac{1}{2\pi}\frac{\gamma}{(\omega-{\varepsilon}
)^2+\gamma^2/4}\,.
\end{align}

\begin{figure}
\begin{center}\includegraphics[width=0.95 \columnwidth]{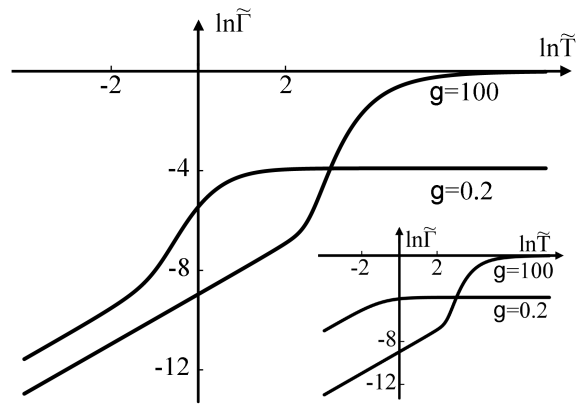}
\caption{Dependence of the decoherence rate on a temperature,
Eq.~(\ref{finiteT}),   for strong and week
 coupling: the main picture is for a fluctuator with
$\nu_{\varepsilon} ({\omega} )$ centered  at
$\widetilde{\varepsilon} =3$, the insert -- for
$\widetilde{\varepsilon} =0$;
  here $\widetilde T,\,
 \widetilde {\Gamma},\,\widetilde{\varepsilon}$  are measured in units of
 ${\gamma} /2$ . }
 \end{center}\end{figure}
The temperature dependence of the decoherence rate (\ref{finiteT})
is presented in Fig.~1 for the strong ($g\!=\!100$) and  weak
($g\!=\!0.2$) coupling to a single fluctuator with
$\nu_{\varepsilon} ({\omega} )$ centered at
$\widetilde{\varepsilon} \equiv 2{\varepsilon} /{\gamma} =3$ (or
 ${\varepsilon} =0$, insert). In the high-$T$ limit, ${\Gamma}$
saturates at the following ${\varepsilon} $-independent value:
 \begin{align}
\Gamma(\infty)=
\frac{\gamma}{2}\left[1-\theta(1-g)\sqrt{1-g^2}\right]\,.\label{noT}
\end{align}
 Although it coincides with the results \cite{paladino,galperin}
of the classical ``high-$T$'' description, it  follows from the
exact expression (\ref{finiteT}) only
 at
$T\!\gg\!\max{\{|\varepsilon_\pm|,\gamma\}} $, where
$\varepsilon_{\pm}\equiv\varepsilon\pm({1}/{2})
\sqrt{v^2-\gamma^2}$. Thus, for a fluctuator with either
relatively large ${\varepsilon} $   or large $v$ (and thus
$|{\varepsilon} _\pm|$), the classical regime  (\ref{noT}) is
never reached. Then ${\Gamma}$ is
  described at any temperature by the ``low-$T$'' asymptotics
   \begin{align}
\Gamma(T)= \frac{T}{\pi}\arctan^2\!\!\left(\frac{2g}{
\widetilde\varepsilon^2 -g^2+1} \right)\,,\label{lowT}
 \end{align}
which follows for any ${\varepsilon} $ from Eq.\ (\ref{finiteT})
\textit{either} for $T\!\ll\!\gamma$ \textit{or} for
$T\ll\min\{|{\varepsilon}_\pm|\}$. This means that the result
(\ref{noT}) of the classical description \cite{paladino,galperin}
is applicable only for a fluctuator with $\nu_{\varepsilon}
({\omega} )$ centered near the Fermi level %of the conduction band
\textit{provided that} $T\gg\max(v,{\gamma} )$.

Note that a crossover between the asymptotics (\ref{lowT}) and
(\ref{noT}) is  relatively sharp when \mbox{$A\equiv \max
({g,\widetilde \varepsilon })\gg1$}: $\Gamma(T)$ changes
exponentially fast,
\begin{align}
\Gamma(T)\sim\Gamma_{\infty}\exp({- {A\gamma}/{2T}}),
\end{align}
in a logarithmically narrow interval, ${A/\ln A}\lesssim
T/{\gamma} \lesssim A$.

 Although a linear in $T$ behaviour similar to that in
Eq.\ (\ref{lowT}) would also follow from the spin-boson models
with the ohmic spectral function, only a full quantum treatment of
a microscopic model, like that in Eq.\ (\ref{htotal}), can result
in a nontrivial $T$-dependence depicted in Fig.~1.

One specific and surprising feature of the model (\ref{htotal}) is
a non-monotonic dependence of the decoherence rate ${\Gamma}$ on
the coupling strength $g$: at low $T$ a contribution of weakly
coupled fluctuators can be orders of magnitude higher than that of
strongly coupled ones, as seen in Fig.~1. Such a non-monotonic
dependence is depicted in Fig.~2 for the fluctuator centered at
$\widetilde{\varepsilon} =3 $. At any finite temperature $\Gamma $
as a function of $g$ has a maximum with a cusp at $g_0=({1+
\widetilde\varepsilon^2 })^{1/2}$. Only at very high $T$
($\widetilde T=100$ in Fig.~2) the cusp is smeared out and
${\Gamma}$ practically saturates at ${\Gamma}(\infty )={\gamma}
/2$ as in the classical limit, Eq.\ (\ref{noT}).

\begin{figure}\begin{center}
\includegraphics[width=0.9\columnwidth]{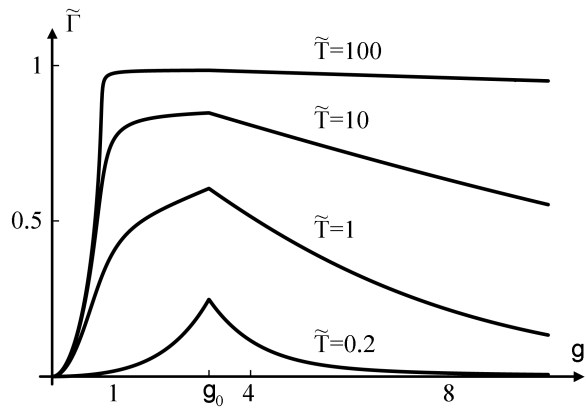}
\end{center}\caption{Non-monotonic dependence of the decoherence rate on the
coupling strength at different temperatures   for
$\widetilde{\varepsilon}=3$ .}
\end{figure}

This surprising suppression   is due to an effective spit in
energy of fluctuators strongly coupled to the qubit. If the qubit
were in one of its eigenstates, the fluctuator energy in the
absence of the hybridization would be split as ${\varepsilon} \pm
v/2$, Eq.~(\ref{htotal}). Allowing for the hybridization, such a
split to the energies
$\varepsilon_{\pm}\equiv\varepsilon\pm({1}/{2})
\sqrt{v^2-\gamma^2}$ occurs only for $g>1$, i.e.\ when the
coupling $v$ exceeds the width ${\gamma} $ of the hybridized
fluctuator Eq.\ (\ref{nu}). Then the decoherence of the qubit in
the mixed state, Eq.\ (\ref{reduced}), is effectively contributed
from two peaks ${\varepsilon} _\pm$ (indeed, the integrand in Eq.\
(\ref{finiteT}) has the two peaks broadened by the hybridization
at ${\omega} ={\varepsilon} _\pm$, besides the exponentially
narrow peak at ${\omega} =0$). Thus,  ${\Gamma}$ increases with
$g$ at ${\varepsilon} =0$ until $v$ reaches ${\gamma} $; a further
increase in $g$ pushes the peaks away from the   Fermi energy,
suppressing the hybridization and thus the switching rate  i.e.\
effectively freezing the charge states. For the arbitrary
${\varepsilon} $, the maximum in ${\Gamma}$ is reached when one of
the peaks at ${\varepsilon} _\pm$ coincides with the Fermi energy.

The results described by Eqs.~(\ref{finiteT}) -- (\ref{lowT}) and
the subsequent discussion refer to the case when there are only a
few fluctuators, so that they can be considered separately and
their decoherence rates could be simply added. If the fluctuators
are dense, one needs to average over the relevant parameters,
which is their energies, coupling constants and switching rates.
The averaging, e.g., over the energy positions ${\varepsilon}_j$,
spread within an interval $E$, would lead to $\Gamma(T)\propto T$
as long as $T\ll E$. The reason is that the number of effective
high-$T$ fluctuators, whose  contribution is described by
Eq.~(\ref{noT}), would be proportional to $T$ while the
contribution of each of low-$T$ fluctuators would  be linear in
$T$, Eq.~(\ref{lowT}).

However, as the decoherence rates due to individual fluctuators
are hugely spread, as illustrated in Fig.~1, the effective
fluctuators are hardly dense as requirements for the effectiveness
are rather restrictive. Firstly, the fluctuator must be not far
from metallic electrodes to be hybridized with conduction
electrons. Secondly, the peaks at $\varepsilon_{\pm}$ should be
within a few $\gamma$'s around the Fermi energy. This brings
further geometrical restrictions essentially reducing the number
of potentially relevant defects so that only relatively few
fluctuators are likely to contribute to decoherence in a typical
experimental setup \cite{nakamura3}.

\acknowledgements We thank B.~L.~Altshuler, R.~Fazio, J.~M.~F.\
Gunn and R.~A. Smith for useful comments. I.V.L.\ acknowledges the
kind  hospitality extended to him at the final stage of this work
at Princeton University and NEC Laboratories America. This work
was supported by the EPSRC grant GR/R95432 and in part by DARPA.

\end{document}